\documentclass{article}

\usepackage{arxiv}

\usepackage[utf8]{inputenc} 
\usepackage[T1]{fontenc}    
\usepackage{hyperref}       
\usepackage{url}            
\usepackage{booktabs}       
\usepackage{amsfonts}       
\usepackage{nicefrac}       
\usepackage{microtype}      
\usepackage{lipsum}
\usepackage{graphicx}
\usepackage{gensymb}

\title{Acoustic Bandgaps of Phononic Penrose and Quasicrystals}

\author{
  Bryan Eastwood \\
  Department of Computer Science\\
  University of Massachusetts Amherst\\
  \texttt{bryan@protonmail.ch} \\
   \And
 Edward A. Rietman \\
  Department of Computer Science\\
  University of Massachusetts Amherst\\
  \texttt{erietman@gmail.com} \\
}

\begin{document}
\maketitle

\fontsize{14}{16}\selectfont

\section{Abstract}
Using 3D printing we manufactured two rodlike phononic crystals. Viewed from the top, one is a Penrose tile and the second is a quasicrystal. We explored the acoustic properties and band gaps for both in the frequency range of 5kHz to 25kHz.


\section{Motivation}
	In our experiment, we measured the acoustic effect on different constructions of solids to the transmission of sound waves. The purpose of this experiment was to corroborate the findings of other researchers on ordered but aperiodic solids on the transmission spectra of waves. Many other experiments have demonstrated a causal relationship between these sort of patterns and the appearance of band-gaps in the spectra, both in the phonoic and photonic domains. Our experiment was designed specifically to determine if this relationship held for our two phononic crystals, the first being structurally based on the Penrose tiling and the second being a quasicrystal. We measured at incidence angles comprising the entire circumference of the crystal to capture the spectra associated with waves bent by its interior constitution.

\section{Procedure}

The design of the two crystals is shown in figure 1. The first is a quasicrystal, which is ordered but not periodic. The second is based off of the Penrose tiling, which is aperiodic. Both crystals have a rod diameter of 6.19mm and a rod spacing of 1.000cm. (3.81mm gap between rods) Both are composed of 3D printed plastic. The experiments were done in air.

\begin{figure}[tbh!]
\centering
\includegraphics[width=0.8\columnwidth]{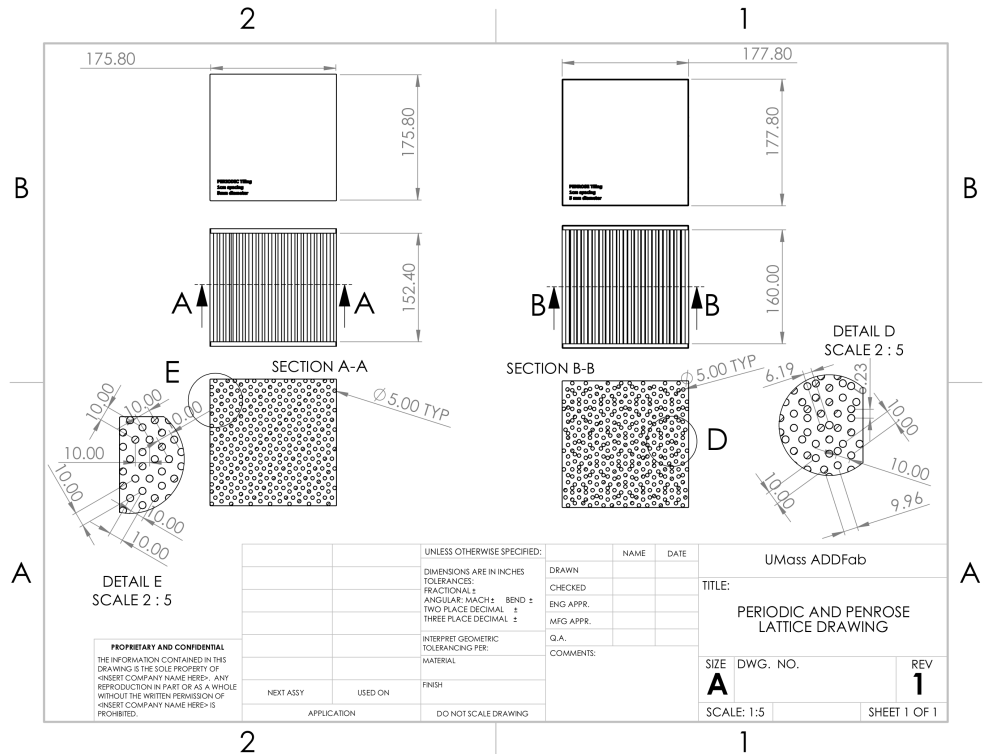}
\caption{construction of the crystals}
\label{crystals}
\end{figure}

	The physical dimensions of the apparatus are shown in figure 2. Each of the microphones is attached to a 16-channel multiplexer, the output of which was connected to an oscilloscope. For each frequency in the range we investigated, a tone of the frequency is played through a speaker. The multiplexer cycles through each of the nine microphones and records the intensity of the frequency using the FFT function of the oscilloscope. Specifically, we record the value of the peak in the frequency spectrum corresponding to the frequency being tested. The frequency spectra shown in this paper are the result of subtracting the results of a trial with each solid placed in the apparatus by the results of a trial with nothing in the apparatus (i.e. the absence of the crystal). This was done because the microphones do not have a flat frequency response.
	
\begin{figure}[h]
\centering
\includegraphics[width=0.8\columnwidth]{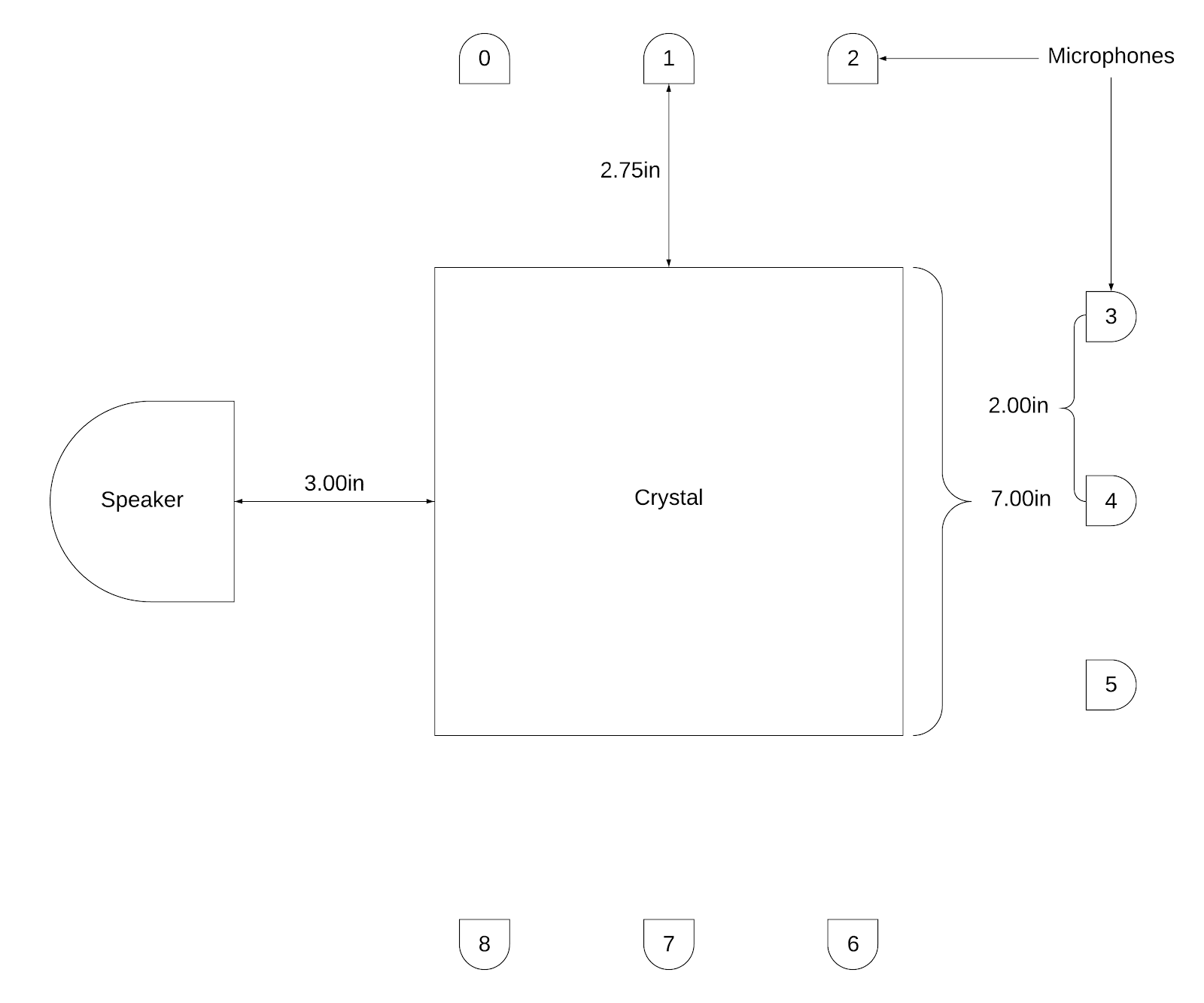}
\caption{ Two-dimensional diagram of the device (top view).
Both the speaker and the microphone are suspended to be level
with the vertical center of the cube (relative to the ground). Numbers
on microphones correspond to the numbering on the plots. }
\label{diagram}
\end{figure}

\section{Results}

	Figure 4 shows the results of the experiment for both solids. Each plot in the 3x3 grid has an abscissa representing our frequency range and an ordinate representing the intensity of the frequency after being picked up by the microphone. We can clearly see that in the central microphone, the band-gap due to the Penrose tiling is wider than that of the quasicrystal, although they are both significant. The spectra from bent waves on either side of the crystal appear to show some kind of periodic behavior, but no band gaps can be discerned.

\begin{figure}[tbh!]
\centering
\includegraphics[width=0.97\columnwidth]{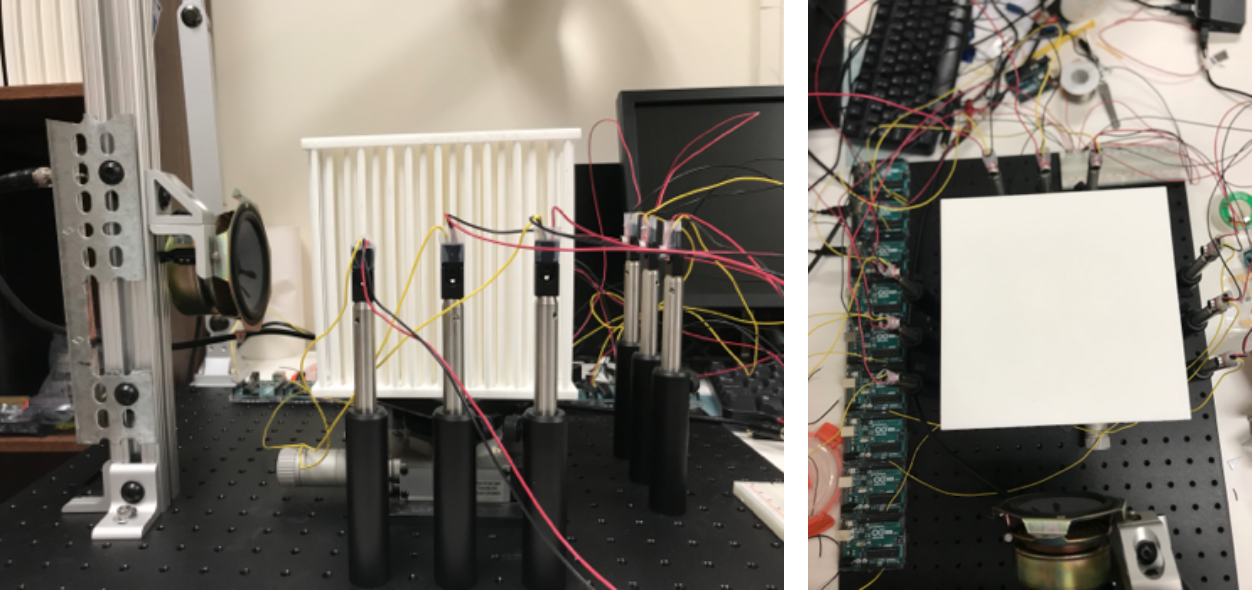}
\caption{ Photos of the layout of the actual experiment.
The wiring shown in these photos was changed to enhance the 
speed of the trials. (Only a single Arduino board was used) }
\label{photos}
\end{figure}

\section{Discussion}

Several other papers have described experiments on the band gaps inherent to both photonic and phononic Penrose\cite{eplphotonic}\cite{prbphotonic} and Quasicrystals.\cite{prbphononic} The main purpose of this experiment was to investigate and attempt to replicate both the theories and experiments of other researchers in this field, and to document our procedure with greater detail. As shown above, we have clearly replicated the presence of band gaps in the transmission spectra of both structures. The isotropy discussed in \cite{prbphononic} was also demonstrated, with a 45\degree{} rotation of the two structures yielding essentially the same result, with the slightly more defined band gap in the 45\degree{} case (figure 5) possibly resulting from the greater amount of crystal between the microphone (due to their orientation and the square shape of the crystals). This experiment could be expanded upon by testing the transmission spectra for much higher frequencies in the order of 100 kHz, in order to replicate the results of the simulations discussed in \cite{rietmanphononic}.

\section{Source Code}

The source code used on the Arduino boards and main computer can be found at \url{https://github.com/bryan-eastwood/phononic-code}.


\bibliographystyle{ieeetr}
\bibliography{bibliography}

\newpage

\begin{figure}[tbh!]
\vspace{0.7cm}
{\centering
\includegraphics[width=\textwidth]{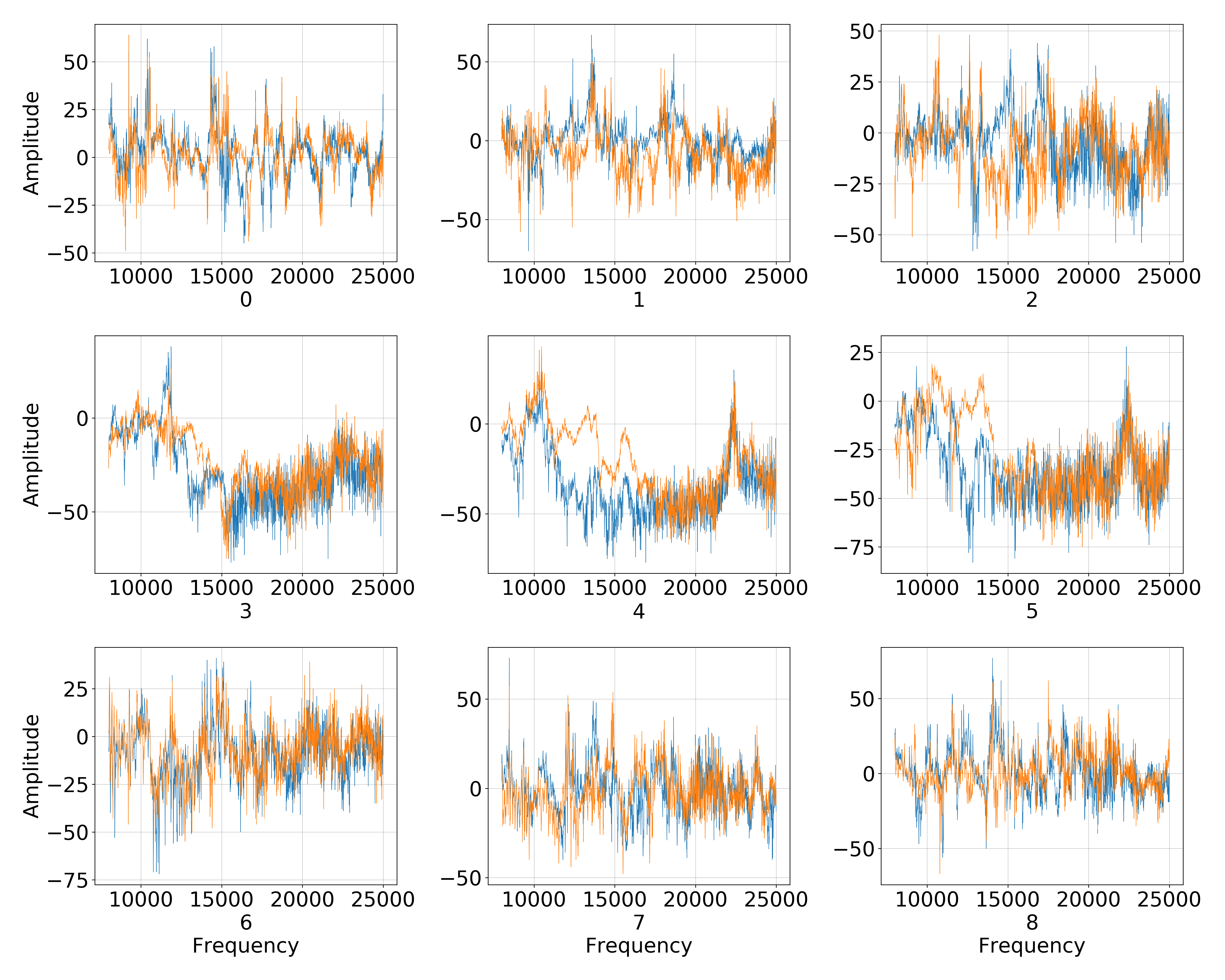}
\caption{ Amplitude vs Frequency (Hz).
Data for phononic crystal shown in blue,
quasicrystal in orange.
 }
\label{fig:alrc}}
\end{figure}

\newpage

\begin{figure}[tbh!]
\vspace{0.7cm}
{\centering
\includegraphics[width=\textwidth]{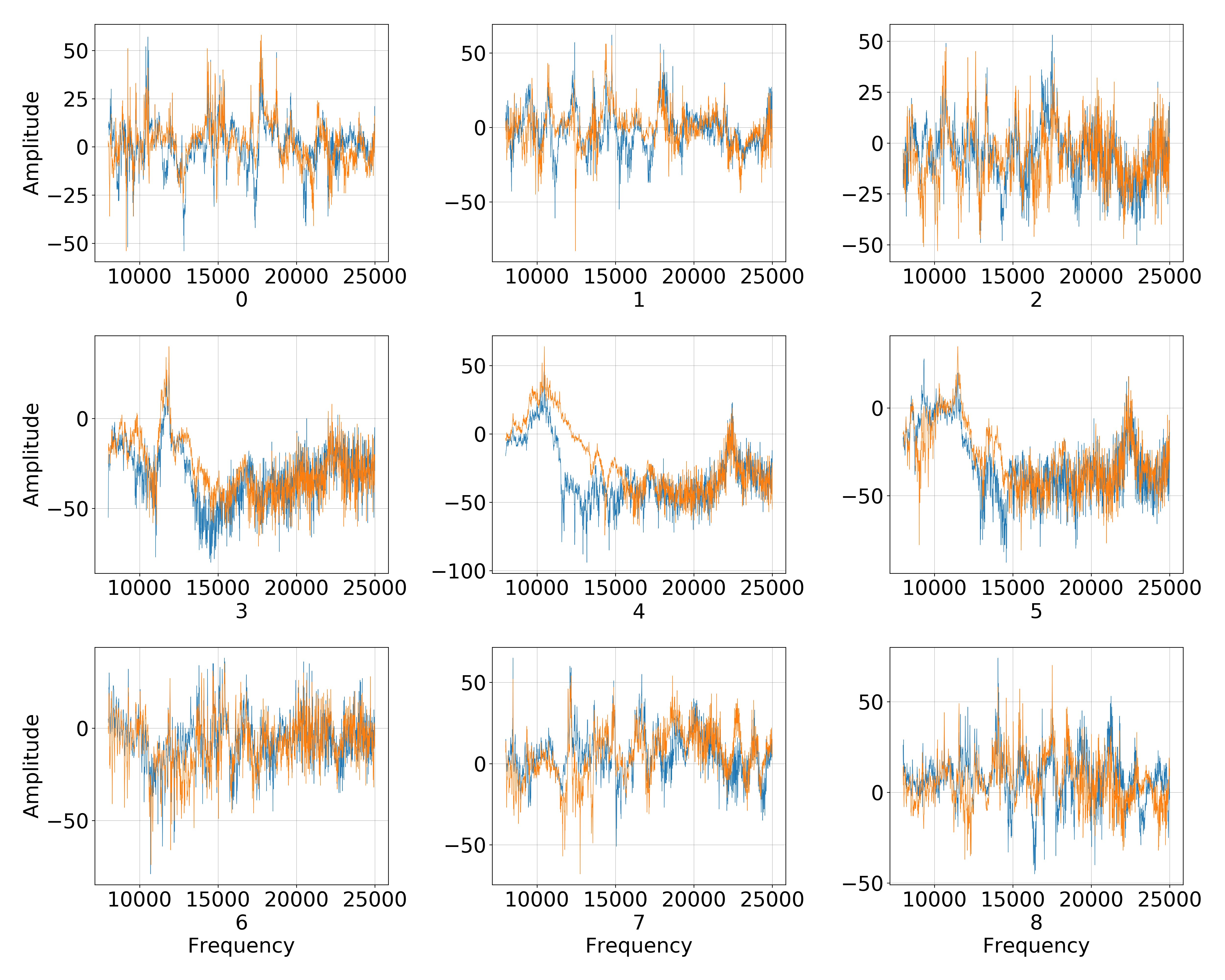}
\caption{ Amplitude vs Frequency at a 45\degree{} rotation (Hz).
Data for phononic crystal shown in blue,
quasicrystal in orange.
 }
\label{fig:alrc}}
\end{figure}

\end{document}